SOFTWARE METAPAPER

# The e-MERLIN Data Reduction Pipeline

Megan K Argo[1]

[1] Jodrell Bank Centre for Astrophysics, University of Manchester, UK

Written in Python and utilising ParselTongue to interface with the Astronomical Image Processing System (AIPS), the e-MERLIN data reduction pipeline is intended to automate the procedures required in processing and calibrating radio astronomy data from the e-MERLIN correlator. Driven by a plain text file of input parameters, the pipeline is modular and can be run in stages by the user, depending on requirements. The software includes options to load raw data, average in time and/or frequency, flag known sources of interference, flag more comprehensively with SERPent, carry out some or all of the calibration procedures (including self-calibration), and image in either normal or wide-field mode. It also optionally produces a number of useful diagnostic plots at various stages so that the quality of the data can be assessed. The software is available for download from the e-MERLIN website or via Github.

**Funding Statement:** The e-MERLIN project has been funded by the University of Manchester, the North West Development Agency, the Science and Technology Facilities Council (STFC), the University of Cambridge and Liverpool John Moores University. Operational funding is provided by STFC and the University of Manchester.

## (1) Overview
### Introduction
Now operational, e-MERLIN [1] is a major upgrade to the Multi-Element Radio-Linked Interferometer Network, MERLIN [2], the United Kingdom's national facility for radio astronomy and Very Long Baseline Interferometry (VLBI). The network consists of five remote antennas operated together with telescopes at Jodrell Bank Observatory (JBO) as an aperture synthesis array. With a maximum baseline length of 217 km, the angular resolution of e-MERLIN at 5-GHz is 40 milli-arcseconds, comparable to that achieved by the Hubble Space Telescope in the visual band. **Table 1** shows the expected final capabilities of e-MERLIN once the upgrades are complete.

The e-MERLIN upgrade has increased the sensitivity of MERLIN by more than an order of magnitude, increased the frequency coverage, and enabled new modes of observation. These improvements are enabling a wide variety of new science, and are a combination of new broadband receivers and associated telescope electronics, together with a dedicated optical fibre network connecting each telescope at a bandwidth of 30 Gb/s to a new flexible high-capacity correlator at JBO. The new correlator, built by DRAO in Penticton, is highly flexible and allows for much more complicated observing modes, and hence a much more diverse range of science, than was possible with the old system.

Compared to MERLIN, datasets from e-MERLIN can be very large, especially when observations are made in 'mixed-mode' with additional high-resolution spectral windows covering astrophysically interesting spectral line transitions. Datasets are pre-processed by e-MERLIN staff using the first few stages of the pipeline described here, and made available to users as UVFITS files with associated tables containing flags and calibration information. Depending on the complexity of the research project and the experience of the user, data may be provided to the user either raw, or partially or fully processed. Options for making data available in other formats (such as CASA measurement sets [3]; ALMA/EVLA SDM and binary data) are under consideration for the longer term.

The staff of the national facility are developing scripts and pipelines to improve the accessibility of data products and images, including an on-line data archive for users to access their data once it has been observed and verified. This paper presents the data reduction pipeline for e-MERLIN, intended to reliably automate the process of data loading, housekeeping, averaging, interference rejection, calibration and imaging. The software is available to all users either from the e-MERLIN website, or via Github.

Briefly, the processes involved in processing a typical e-MERLIN continuum dataset are as follows. Each of these steps can be carried out by the pipeline either individually, or all of them in sequence, by varying the input parameters. Firstly the raw data (exported from the local archive in FITS-IDI format after correlation) must be loaded into AIPS. At this point a typical observation consists of multiple files, at least one per observed astronomical object.

| | 1.5 GHz (L-band) | 5 GHz (C-band) | 22 GHz[8] (K-band) |
|---|---|---|---|
| Resolution[1] (mas) | 150 | 40 | 12 |
| Field of view[2] (arcmin) | 30 | 7 | 2 |
| Frequency range (GHz) | 1.3 – 1.7 | 4 – 8 | 22 – 24 |
| Bandwidth[3] (GHz) | 0.4 | 2 | 2 |
| Sensitivity[4] (μJy/bm) | 5 – 6 | 1.8 – 2.3 | ~15 |
| Surface brightness sensitivity[4] (K) | ~190 | ~70 | ~530 |
| Astrometric performance[5] (mas) (ICRF) | ~2 | ~1 | ~2 |
| Astrometric performance[6] (mas) (day-to-day) | ~0.5 | ~0.2 | ~1 |
| Amplitude calibration[7] (%) | 2 | 1 | 10 |

*Notes* 1: with uniform weighting, at the central frequency. 2: FWHM of 25-m dishes; reduced when the Lovell telescope is included. 3: Maximum bandwidth per polarization. 4: In a full imaging run with the Lovell telescope. 5: With respect to the ICRF (typical 3-deg target-calibrator separation using the VLBA Calibrator Survey). 6: Day-to-day repeatability using surveyed or in-beam sources, and assuming full imaging run. 7: Targets for day-to-day repeatability. 8: The Lovell telescope is not available at 22 GHz.

**Table 1:** Technical capabilities of e-MERLIN once the upgrade is complete. The Lovell telescope may be included in the array at 1.5 and 5 GHz; this increases the sensitivity by a factor of between 2 and 3, but reduces the field of view. For current capabilities, see http://www.e-merlin.ac.uk/

The data are then sorted if required, and some other housekeeping procedures are carried out. Once the data are loaded, a flag mask may be applied; this step creates a flag table for each dataset using a list of frequencies where interference is known to affect e-MERLIN telescopes. Averaging in time and/or frequency is then carried out, with the amount of averaging determined by the user depending on the needs of the experiment. If required, the data may be sorted at this point. The pipeline then produces a series of pre-calibration diagnostic plots which can be used to assess the quality of the data before proceeding with calibration. Automated flagging with SERPent [4] is then carried out, removing further interference-affected data, and post-flagging diagnostic plots may also be produced. At this point the averaged, flagged data may be backed up to disk outside of AIPS. The next step concatenates together all of the individual datasets, producing one dataset which can be calibrated. The data are then calibrated using standard procedures for interferometric arrays, including: fringe fitting, flux calibration, calculation of bandpass solutions, phase calibration, and optional self-calibration. Imaging may then be carried out, either as a single field, or in wide-field mode where the dataset is split up into small chunks, phase-rotated, imaged, and then mosaicked together. Any or all of these steps may, of course, be carried out by hand by the user without making use of the pipeline. A detailed description of the process of e-MERLIN data reduction covering all possibilities (including recommendations for dealing with mixed-mode spectroscopic data, polarisation calibration, astrometric observations, etc.) is beyond the scope of this paper since they are not yet implemented in the pipeline, the recommendations for dealing with these modes are still evolving, and current guidelines are provided in the e-MERLIN Cookbook [5] which is regularly updated.

Papers resulting from e-MERLIN datasets which have so far passed through the pipeline as part of the testing process include [6], [7], [8] and [9]. A large part of the development was carried out using the data published in [6], and data published in [7], [8] and [9] was also used to directly test pre-release versions of the code. Other publications which have resulted from data passed through development versions of the pipeline include: [10] [11] [12] and [13].

**Implementation and architecture**
Written in Python, the pipeline uses ParselTongue [14] to interface with AIPS [15], one of the main standard software packages for radio astronomy data reduction. The pipeline distribution consists of several files which are required for operation. The pipeline itself is contained in '*eMERLIN_pipeline.py*', and requires '*eMERLIN_tasks.py*' which contains many task definitions. A plain text file of inputs is also required; this can be named anything but must be specified on the command line at runtime. **Table 2** shows the main control parameters, used to select which tasks to perform. Many of these tasks have additional parameters, all of which are documented in the example file '*doall.inputs*' which is contained in the pipeline distribution. Any of the steps can be performed by setting the appropriate parameter to equal 1. The order of the steps is such that a raw dataset can be fully processed, in the correct order (described above), by simply telling the script to run everything.

Depending on which tasks are to be run, additional files may also be required. When applying the flag mask to flag (blank out) certain frequencies based on known sources of interference seen at e-MERLIN antennas, the file '*flagmask512.fg*' must be in the working directory. Likewise, if applying the SERPent autoflagger [4], '*SERPent.py*' must be present. The user must have write permissions in the current directory, and (if applicable) also in the output locations specified for fits backups and diagnostic plots.

| | |
|---|---|
| userno = 19 | AIPS number to use (no default) |
| indisk = 1 | AIPS disk to use |
| targets = B1938+666 | List of target sources |
| phsrefs = J1939+123 | List of phase calibrators |
| fluxcals = 1331+305 | List of flux calibrators |
| bpasscals = 1407+284 | List of bandpass calibrators |
| pointcals = 1407+284 | Source to use as point calibrator |
| widetargets = B1938+666 | List of targets for wide-field imaging |
| fitsdir = /local/scratch/user/ | Location of data to load |
| plotdir = /local/scratch/user/ | Where to write plots, defaults to current working directory |
| fittpdir = /local/scratch/user/ | Where to backup fits, default as above |
| doload * | Load data (from list in *fitsdir*) |
| doflagmask | Apply the e-MERLIN flagmask |
| doavg * | Average data in time and/or frequency |
| dodiagnostic1 | Make diagnostic plots |
| doflag * | Apply SERPent autoflagger |
| dodiagnostic2 | Post-flagging diagnostic plots |
| doback | Backup the flag tables |
| doconcat | Concatenate data into one database for calibration |
| docalib * | Perform some or all of the calibration procedures |
| doimage | Image the *phsrefs* and *targets* |
| dowide * | Image the full primary beam for the sources in *widetargets* |

**Table 2:** Input parameters required for the operation of the e-MERLIN pipeline. All of the listed tasks are performed if set equal to 1, and not performed if set to -1. If not present in the inputs file, they all default to -1 in order to avoid damaging data already present in the user's AIPS catalogue. Tasks denoted with * are those which require additional parameters (documented in the example "doall.inputs" which is distributed with the pipeline).

The pipeline is executed by running > *parseltongue pipeline.py inputs.txt* on the command line, where *inputs.txt* is a plain text file containing a list of inputs necessary to tell the pipeline where the data are located and what steps are required (see **Table 2**). When downloaded from the repository, the pipeline comes with an example inputs file ('*doall.inputs*') which includes comments describing each parameter. In most cases, if a parameter is missing from the inputs file, a sensible default is used. When executed, the inputs are printed to the terminal window, along with a listing of the specified AIPS catalogue, before any tasks are performed, and the user is asked to check the inputs and confirm whether or not to continue.

Parameters which can be calculated from the data (such as the pixel size for imaging) can also be specified to override the numbers calculated in the pipeline. Sanity checks are made on some parameters to ensure that the numbers are sensible (a useful pixel size for example, or that the specified reference antenna is actually present in the dataset), or that the file catalogue within AIPS is in the expected state compared to the tasks requested.

### Control parameters
The operation of the pipeline is entirely driven by the inputs file. The main parameters are listed in **Table 2**, and further details are provided in the e-MERLIN Cookbook [5].

### Input
The input depends on which tasks are being applied. Datasets directly from the e-MERLIN online archive consist of several files, written in the standard FITS-IDI format used widely in astronomy. Each file consists of correlated visibilities for a single source observed during the project. The first stage of the pipeline (*doload*) expects raw data from the correlator as multiple single-source fits files on disk outside of AIPS. The next stages of the pre-calibration process (*doflagmask, doavg, dodiagnostic1, doflag, dodiagnostic2, doback, doconcat*) expect one or more single-source files already loaded into the specified AIPS user catalogue. All sub-processes within the calibration section (*docalib*) expect a single multi-source file, and the two optional imaging routines (*doimage, dowide*) expect a multi-source file with appropriate calibration tables attached. From calibration onwards, if only one file is present in the AIPS catalogue specified then it is assumed that this file should be used; if more than one file is present, the pipeline expects to find one file named 'ALL. DBCON.1', the name given to the output from the concatenation stage. If neither of these conditions are matched, execution is terminated with an explanatory message.

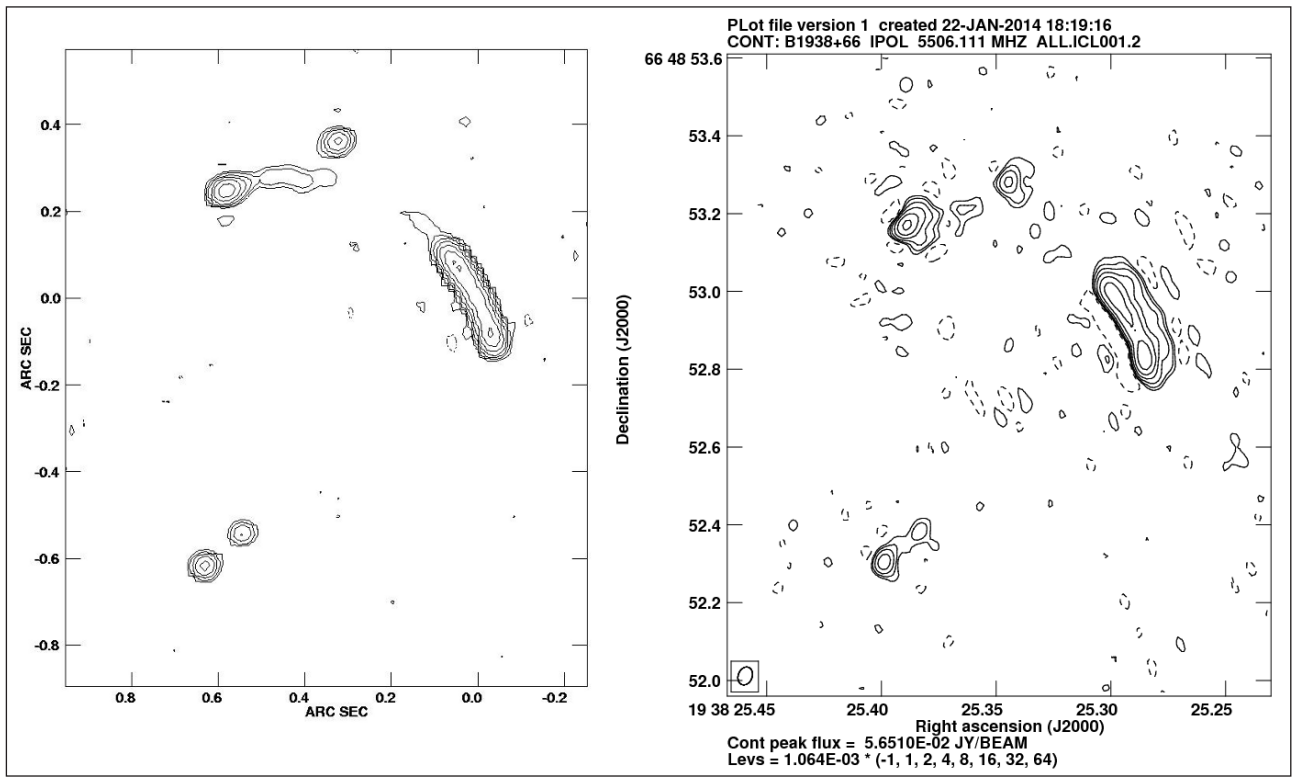

**Figure 1:** Left: archival C-band image of the gravitational lens system B1938+666 from MERLIN data. Right: pipelined image of the same source, made using the C-band verification dataset and v0.7 of the pipeline with no human intervention.

### Output
Again, the output depends on which stage(s) of the pipeline are being run. The first stage (*doload*) results in the files loaded into the AIPS catalogue. The averaging step (*doavg*) averages each file and deletes the unaveraged data from the AIPS catalogue to free up space (the original data files are never deleted from the file-system by the pipeline). After concatenation (*doconcat*) all of the files in the catalogue will have been combined into one UV database within AIPS. The imaging tasks (*doimage, dowide*) result in cleaned images of the requested sources. Other tasks (*dodiagnostic, doflagmask, doflag, docalib*) result in calibration or flag tables appended to the files which have been operated on.

### Quality control
Extensive testing of the pipeline has been carried out by competent staff using a variety of production e-MERLIN datasets. The loading, housekeeping, flag mask and concatenation sections of the pipeline work reliably and are now used on every observed project. For well-behaved datasets, those without serious problems during either observation or correlation, the calibration section of the pipeline will produce useful data products. However the complexities of an operational interferometer and the nature of calibration are such that a suitably-trained human will always do a better job than an unintelligent piece of software; users are advised to check carefully the data products from this stage of the pipeline. Testing of the pipeline has, in fact, brought to light some interesting problems occurring further upstream in the e-MERLIN system.

For verification purposes, the main test datasets from v0.7 onwards are two observations carried out for use at a data reduction school held at JBCA in Manchester in January 2014. These datasets both consist of standard continuum phase-referencing observations of the gravitational lens system B1938+666, observed at frequencies of 1.4 and 5 GHz. Both datasets have been carefully processed by a human so that the data are well-understood, and reference data products are available to provide a direct comparison to the automated pipeline output. **Figure 1** shows a pipeline-processed (with v0.7) image of the C-band verification dataset, together with an archive image of the same source produced by a human as part of a radio survey. For training and testing purposes, both of these verification datasets are available on the e-MERLIN website for download [5]. These two datasets will be run through each new release of the pipeline for verification, before the code is publicly released on Github.

## (2) Availability
### Operating system
The pipeline will run on any linux system capable of running the underlying packages: AIPS (31DEC13 or later) [15], Obit [16], and Parseltongue [14].

**Programming language**
Python 2.7

**Additional system requirements**
If running the complete pipeline from raw correlator output, sufficient disk space is required to cope with twice the size of the input dataset (since AIPS essentially copies the data on loading), plus additional space to hold the averaged datasets. If data has been pre-processed by e-MERLIN staff, much less disk space is required, but performance may still be limited by disk I/O capacity.

**Dependencies**
Python 2.7; Obit; AIPS 31DEC13 or later; ParselTongue 2.0

**List of contributors**
The following have all provided segments of code: Neal Jackson, Rob Beswick, Nick Wrigley, Pierre-Emmanuel Belles (JBCA/e-MERLIN); Danielle Fenech (UCL)

**Support**
User support for e-MERLIN, including use of the pipeline, is available from e-MERLIN staff located at Jodrell Bank Centre for Astrophysics (e-merlin@jb.man.ac.uk).

**Software location**
*Archive*
(e.g. institutional repository, general repository) (required)

*Name*
e-MERLIN website

*Location*
http://www.e-merlin.ac.uk/observe/pipeline/

*Persistent identifier*
ascl: 1407.017

*Licence*
GPLv3

*Publisher*
Megan Argo

*Date published*
27/05/2014 (v0.7)

**Code repository**
(e.g. SourceForge, GitHub etc.) (optional)

*Name*
GitHub

*Identifier*
http://github.com/mkargo/pipeline/

*Persistent identifier*
doi: 10.5281/zenodo.10163 (v0.7)

*Licence*
GPLv3

*Date published*
27/05/2014 (v0.7)

**Language**
The pipeline is written in Python using ParselTongue, the python interface to AIPS.

## (3) Reuse potential

This software is intended for both local facility support staff and external facility users. All data from the e-MERLIN correlator requires certain initial processing steps which are included in the first few stages of the pipeline, allowing facility staff to pre-process data rapidly and examine the data quality of an observation prior to releasing the data to users. Users of the instrument may then calibrate and image their data from scratch entirely by hand, or use the pipeline if they choose.

The software is deliberately designed to be run in stages. This is partly so that users can choose carry out only certain steps, inspecting the data and calibration solutions in between each step, and partly so that new sections of code are simple to add. The section dealing with interference flagging via SERPent, and that designed to carry out wide-field imaging, were both added to the code after much of the initial development had been carried out.

As more sophisticated modes of observation are tested and become available to users (wide-field, mosaicking, mixed-mode spectral line, polarimetry, joint observations with other facilities, etc.), new sections of code to deal with these modes can be added as new modules in the same way. This is particularly important for the teams working on the large e-MERLIN Legacy programmes, all of which involve several hundred hours of observations. Each of the Legacy projects has specific data reduction needs particular to their science goals, and the pipeline can be easily modified to accommodate these requirements.

The next developments to be implemented will deal with cases where multiple source-phase calibrator pairs exist, where mixed-mode spectral line observations include sub-bands of different widths (something which AIPS was not designed to cope with), implementing polarisation calibration, and refining the flux calibration using observed models of 3C286, the standard e-MERLIN calibrator.

In the longer term, once the newer data reduction package CASA [3] becomes capable of calibrating data from the e-MERLIN correlator, the entire calibration process can migrate from AIPS to CASA. Since CASA is already capable of interfacing with Python, it should be reasonably straightforward to convert the pipeline for use with CASA.

**Acknowledgements**
Thanks to the early users of e-MERLIN who tested various versions of the script, provided valuable feedback and bug reports, or allowed me to use their datasets for verification purposes. Thanks also to the authors and maintainers of the EVN pipeline, which provided the inspiration and a starting point for this work.

**References**

Codes which are listed in the Astrophysics Source Code Library (ascl.net) have their ascl identifiers listed in the appropriate reference below.

1. **Spencer, R.** Progress and Status of e-MERLIN. In proceedings of the 8th International e-VLBI Workshop, Madrid, (2009). http://adsabs.harvard.edu/abs/2009evlb.confE..29S
2. **Thomasson, P.** MERLIN. QJRAS 27 413 (1986). http://adsabs.harvard.edu/abs/1986QJRAS..27..413T
3. CASA is available from: http://casa.nrao.edu/ [ascl:1107.013]
4. **Peck, L., Fenech, D.** SERPent: Automated reduction and RFI-mitigation software for e-MERLIN. Astronomy & Computing 2 54 (2013). DOI: http://dx.doi.org/10.1016/j.ascom.2013.09.001 Code: [ascl:1312.001]
5. The current version of the e-MERLIN Cookbook is available from: http://www.e-merlin.ac.uk/data_red/ (accessed May 30th 2014).
6. **Argo, M. K., Paragi, Z., Röttgering, H., Klöckner, H.-R., Miley, G., Mahmud, M.** 2013 Probing the nature of compact ultra-steep spectrum radio sources with the e-EVN and e-MERLIN. MNRAS Letters 431 L58 (2013). DOI: http://dx.doi.org/10.1093/mnrasl/slt008
7. **Romero-Cañizales, C., Pérez-Torres, M. A., Alberdi, A., Argo, M. K., Beswick, R. J., Kankare, E., Batejat, F., Mattila, S., Conway, J. E., Efstathiou, A., Garrington, S. T., Muxlow, T. W. B., Ryder, S. D., Väisänen, P.** 2012 e-MERLIN and VLBI observations of the luminous infrared galaxy IC 883: a nuclear starburst and an AGN candidate revealed. A&A 543 72 (2012). DOI: http://dx.doi.org/10.1051/0004-6361/201218816
8. **Perez-Torres, M., Argo, M., Lundqvist, P., Anderson, G., Beswick, R., Bjornsson, C. I., Fender, R., Rushton, A., Ryder, S., Staley, T.** 2013 5.0 GHz Continuum MERLIN Observations of the Type Ia SN 2013dy. ATel #5619 (2013). http://www.astronomerstelegram.org/?read=5619
9. **Pérez-Torres, M. A., Lundqvist, P. Beswick, R., Bjornsson, C. I., Muxlow, T. W. M., Paragi, Z., Ryder, S., Alberdi, A., Fransson, C., Marcaide, J. M., Marti-Vidal, I., Ros, E., Argo, M. K., Guirado, J. C.** 2014 Constraints on the progenitor system and the environs of SN 2014J from deep eMERLIN and EVN observations. ApJ 792 38 (2014).
10. **Guidetti, D., Bondi, M., Prandoni, I., Beswick, R. J., Muxlow, T. W. B., Wrigley, N., Smail, I., McHardy, I.** 2013 e-MERLIN observations at 5 GHz of the GOODS-N region: pinpointing AGN cores in high-redshift galaxies. MNRAS 432 2798 (2013). DOI: http://dx.doi.org/10.1093/mnras/stt633
11. **An, T., Paragi, Z., Frey, S., Xiao, T., Baan, W. A., Komossa, S., Gabanyi, K. Ã., Xu, Y.-H., Hong, X.-Y.** The radio structure of 3C 316, a galaxy with double-peaked narrow optical emission lines. MNRAS 433 1161 (2013). DOI: http://dx.doi.org/10.1093/mnras/stt801
12. **Richards, A. M. S., Davis, R. J., Decin, L., Etoka, S., Harper, G. M., Lim, J. J., Garrington, S. T., Gray, M. D., McDonald, I., O'Gorman, E., Wittkowski, M.** e-MERLIN resolves Betelgeuse at $\lambda$ 5 cm: hotspots at 5 R. MNRAS 432 61 (2013). DOI: http://dx.doi.org/10.1093/mnrasl/slt036
13. **Gendre, M. A., Fenech, D. M., Beswick, R. J., Muxlow, T. W. B., Argo, M. K.** Flux density variations of radio sources in M82 over the last three decades. MNRAS 431 1107 (2013). DOI: http://dx.doi.org/10.1093/mnras/stt231
14. **Kettenis, M., Sipior, M.** ParselTongue: AIPS Python Interface. Astrophysics Source Code Library (2012). http://www.jive.nl/wiki/doku.php?id=parseltongue:parseltongue [ascl:1208.020]
15. AIPS is available from: http://www.aips.nrao.edu/ [ascl:9911:003]
16. **Cotton, W. D.** Obit: A Development Environment for Astronomical Algorithms. PASP 120 866 (2008). DOI: http://dx.doi.org/10.1086/586754 [ascl:1307.008]